\documentclass{mem}
\usepackage{natbib}\usepackage{txfonts}\usepackage{balance}
\usepackage{graphicx}
\usepackage[a4paper,breaklinks,dvipdfm]{hyperref}
\idline{84}{1}
\usepackage{txfonts} 

\begin{document}
\def\teff{$T\rm_{eff }$}
\def\kms{$\mathrm {km s}^{-1}$}

\title{
Tackling the soft X-ray excess in AGN with variability studies
}

   \subtitle{}

\author{
A.\,M.\,Lohfink\inst{1}, C.\,S.\,Reynolds\inst{1}, R.\,F.\,Mushotzky\inst{1},  
\and M.\,A.\,Nowak\inst{2}
          }

  \offprints{A.\,Lohfink}

\institute{
Department of Astronomy
University of Maryland
College Park, MD 20742-2421, USA
\and
Massachusetts Institute of Technology 
Kavli Institute for Astrophysics
Cambridge, MA 02139, USA\\
\email{alohfink@astro.umd.edu}
}

\authorrunning{Lohfink }

\titlerunning{Soft Excess Variability}

\abstract{
The origin of the soft X-ray excess in AGN has been a mystery ever since its discovery. We present how the time variability of this spectral component can point towards its origin. Using the powerful technique of multi-epoch fitting, we study how the soft excess in a given object depends on other parameters of the continuum and the accretion disk possibly hinting at its nature. As an example, we present results from this technique applied to the Seyfert galaxy Mrk 841. We study all (3) XMM and some of the Suzaku pointings available and find that the source displays an impressive variability in the soft X-ray band on the timescale of years. We study several common soft excess models and their ability to physically consistently explain this spectral variability. Mrk\,841 is found to show a distinct variability pattern that can be best explained by the soft excess originating mostly from a thermal Comptonization component. The variability timescale can be constrained to be on the order of a few days. 
}
\maketitle{}

\section{Introduction}
The X-ray spectrum from most Seyfert galaxies can be characterized by a power law continuum with reflection from the accretion disk and the torus. However, there is one portion of the X-ray spectrum that is mysterious even after a decade of \textit{XMM} and \textit{Suzaku} observations: the soft X-ray excess (``soft excess" hereafter), i.e. the excess flux at soft X-ray energies with regard to a power law continuum. This feature is common in active galactic nuclei (AGN) X-ray spectra but its physical nature is completely uncertain.  

The spectral models able to describe the soft excess are highly degenerate \citep{page:04a,lohfink:12b} and even high resolution spectroscopy obtained with \textit{Chandra} and XMM has not yielded any additional insights or been able to break the degeneracies \citep{turner:00a}. Due to these modeling difficulties, the physical origin of this soft excess is highly uncertain. Understanding its nature is crucial because of its potentially large luminosity (depending on its exact shape) and the influence it has on the detailed spectral shape of the continuum. For example, uncertainty in the soft excess shape can be an important source of systematic error for the spin parameter of the black hole (see recent analysis by \citealt{nardini:11a} and \citealt{lohfink:12b}). 

\begin{figure}[t!]
\includegraphics[width=\columnwidth]{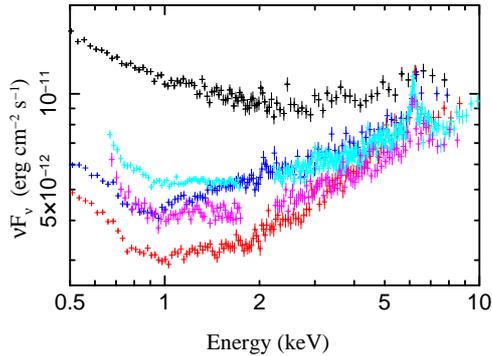}
\caption{\footnotesize
Unfolded spectra (\textit{Suzaku} key programme [purple, turquoise] \& \textit{XMM} [black, blue,red]) of Fairall~9 showing the spectral variability of the source. The spectra were rebinned for plotting.}
\label{eta}
\end{figure}

Ever since its discovery over 25\,years ago \citep{singh:85a}, there has been a debate about what is producing this excess and there are many distinct ideas as to its physical nature. While the currently two most popular ideas are blurred ionized reflection from the inner parts of the accretion disk \citep{gierlinski:04a,crummy:06a}, and Comptonization components \citep{ross:92a} many more have been proposed. For radio-loud objects and narrow line Seyferts an additional power law or broken power law can describe the soft excess rather well \citep{kataoka:07a}. Physically this power law component can be associated with an optically thick Comptonization component \citep{papadakis:10a} or a jet component \citep{chatterjee:09a}.  

A variety of methods have been used to determine the nature of the soft excess is and have narrowed down the options to only two possibilities. Moreover the study of AGN samples and multi-wavelength studies have led to further insights into the properties of the soft excess. From the CAIXA sample we learned that no correlation of strength of the soft excess with black hole mass or luminosity of the AGN exists \citep{bianchi:09a}. Multiwavelength studies have revealed a possible dependency of the UV slope with the soft excess strength and shape \citep{walter:93a,atlee:09a}. While this is valuable information, to make further progress regarding the soft excess a new approach is needed. An analysis of the luminous Seyfert~1 galaxy Fairall~9 by \citet{lohfink:12b} already emphasized that the soft excess is variable. Assembling the most recent \textit{Suzaku} key programme spectra of the Seyfert~1 galaxy Mrk\,841 together with all the archival \textit{XMM} data available, we confirm this variability (Fig.~\ref{eta}) also in Mrk\,841. While individual objects have been studied in the past, it is important to note that these mostly have been studies of single epoch pointings, ignoring the variability. 

\section{Spectral Analysis}
The major unresolved question regarding the soft excess is whether the soft excess is characterized by a separate spectral component (e.g. \citealt{noda:12a}) or is part of some broad band component, like ionized reflection. In order to address this question and eventually determine the physical nature of the soft excess, a new approach is needed. Here we utilize the time variability of the soft excess to test which model leads to the best description. 

From previous analyses it is clear that the X-ray spectrum of Mrk\,841 can be well described by a 2-zone warm absorber, a continuum (modeled as a power law), cold and ionized reflection, and a soft excess. For the soft excess we consider three different possibilities: a) the soft excess is caused by the blurred ionized reflection in the spectrum, b) the soft excess is a superposition of multiple ionized layers of the accretion disk or c) the soft excess is an additional thermal Comptonization component. The different ideas were tested performing a multi-epoch fit to all the datasets shown in Fig.~\ref{eta}, where the key parameters such as spin, accretion disk inclination and iron abundance are tied during the fitting. A single ionized reflector leads to an unacceptable fit and will not be further discussed. The remaining two models lead to somewhat comparable fit qualities with the Comptonization being statistically preferred.

\subsection{Compton Model}
\begin{figure*}[h!tb]
\begin{minipage}{0.45\textwidth}
\includegraphics[width=\textwidth]{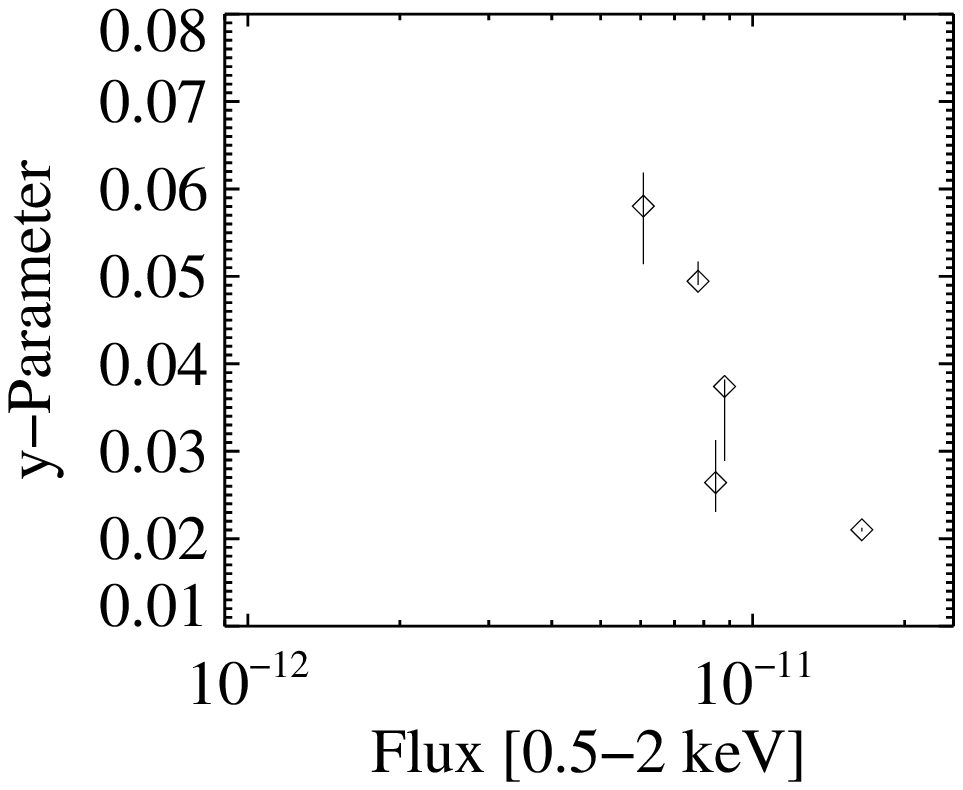}
\caption{\footnotesize Soft X-ray flux versus Compton-$y$ parameter of the soft excess. A correlation is apparent.}\label{comp1}
\end{minipage}
\hfill
\begin{minipage}{0.45\textwidth}
\includegraphics[width=\textwidth]{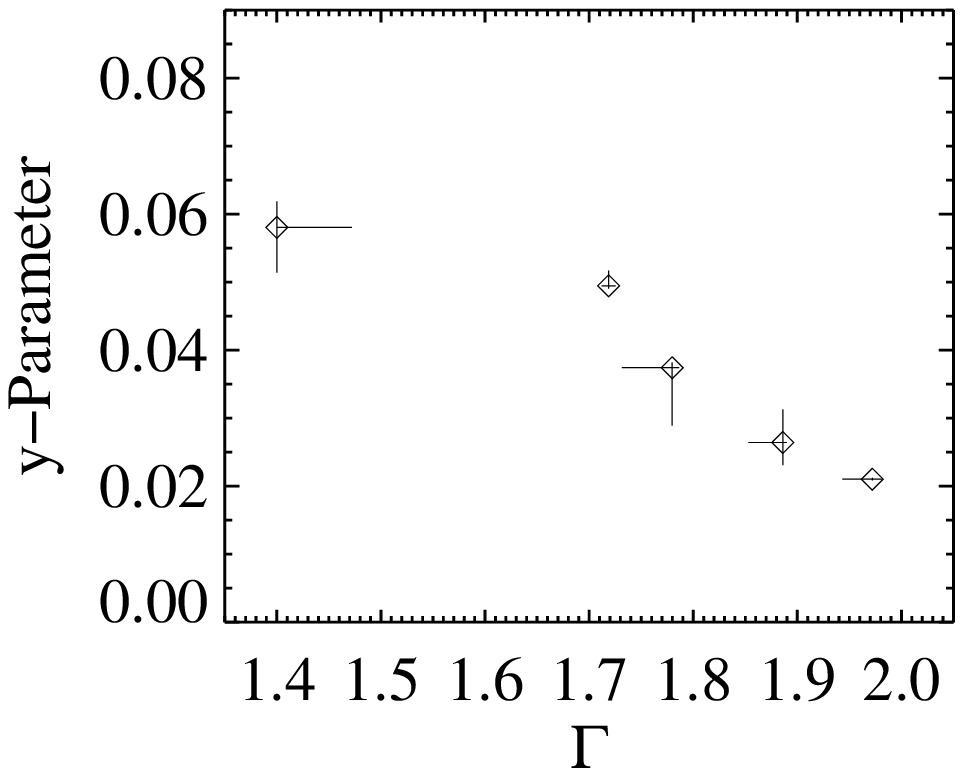}
\caption{\footnotesize Photon Index ($\Gamma$) versus Compton-$y$ parameter of the soft excess. A correlation is apparent.}\label{comp2}
\end{minipage}
\end{figure*}

For a fit including an additional Comptonization component at soft energies to model the soft excess only the Compton-$y$ parameter can be constrained not $kT$ and $\tau$ separately. We observe a correlation between the soft X-ray flux and the Compton-$y$ parameter (Fig.~\ref{comp1}). The more soft X-ray flux the smaller the Compton-$y$. Another correlation exists between the hard X-ray photon index and the Compton-$y$ of the soft Compton component (Fig.~\ref{comp2}). The overall spectrum becomes steeper as the Compton-$y$ parameter decreases. 

\subsection{Reflection Model} 
\begin{figure*}[h!tb]
\begin{minipage}{0.45\textwidth}
\includegraphics[width=\textwidth]{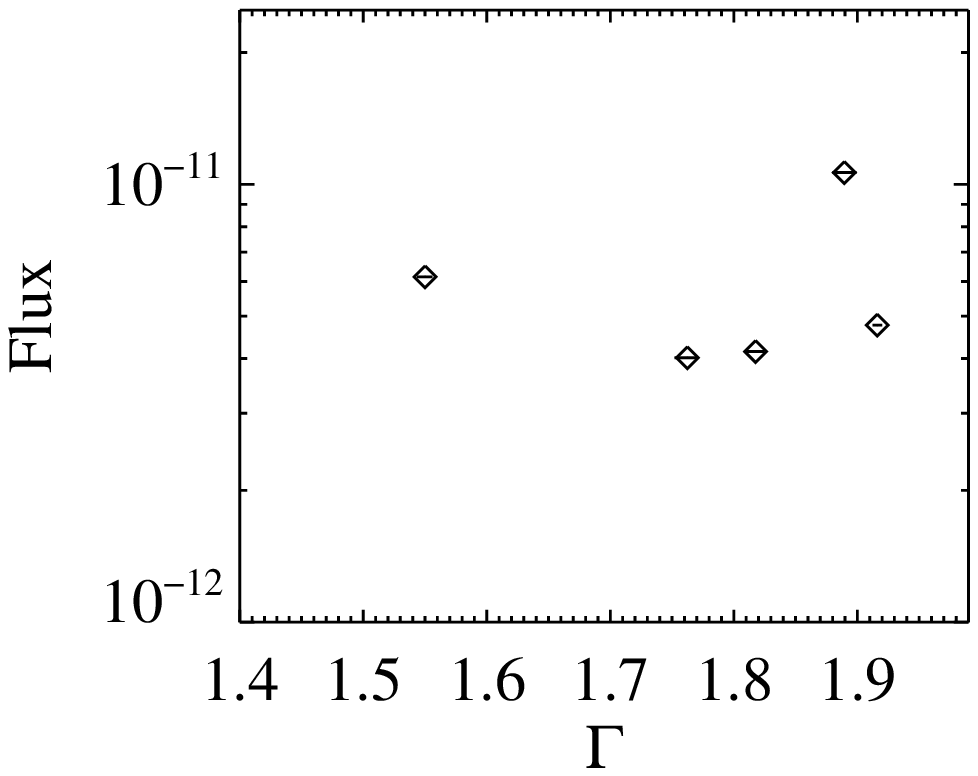}
\caption{\footnotesize Photon Index ($\Gamma$) versus soft X-ray flux. No correlation is apparent in the reflection case.}\label{refl1}
\vspace*{1.2\baselineskip}
\end{minipage}
\hfill
\begin{minipage}{0.45\textwidth}
\includegraphics[width=\textwidth]{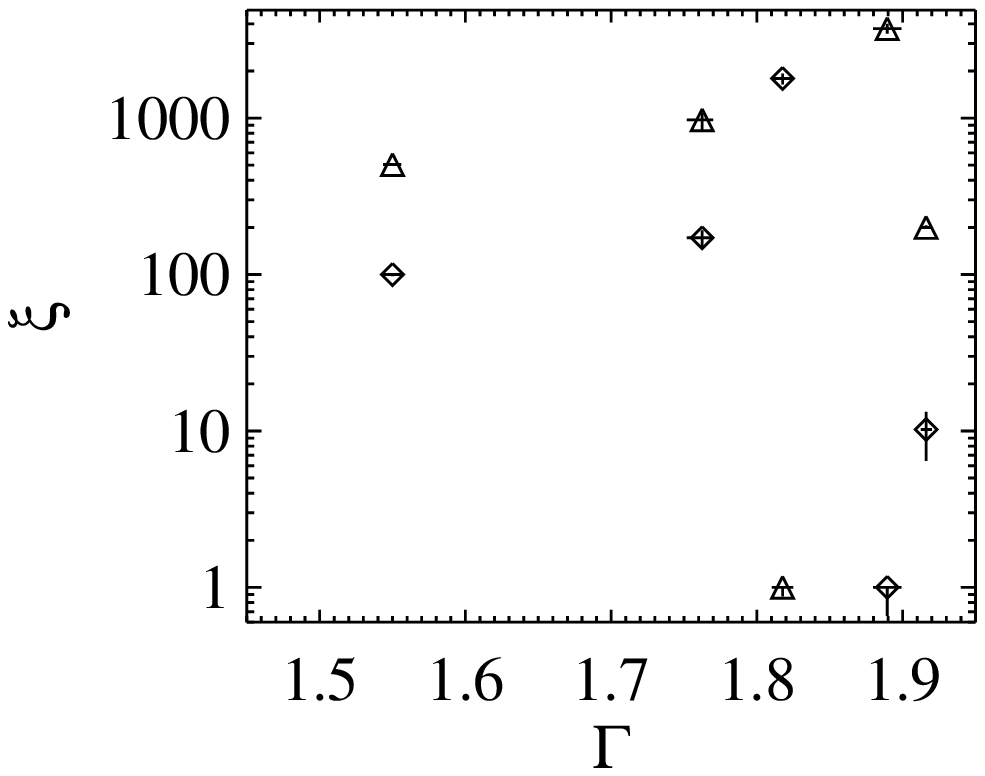}
\caption{\footnotesize Photon Index ($\Gamma$) versus inner (triangle) and outer (diamond) ionization parameter. No correlation is apparent.}\label{refl2}
\end{minipage}
\end{figure*}

Based on the correlations seen for a model including a soft Compton component one would expect to also see correlations for a model with just blurred ionized reflection. Contrary to the expectations however there is no correlation between the soft X-ray flux and the photon index of the overall spectrum (Fig.~\ref{refl1}). Moreover, one would expect a correlation between the ionization states of the reflectors which is not the case either (Fig.~\ref{refl2}). In fact at times the ``outer" reflector is more ionized than the ``inner" reflector pointing towards clear problems in the modeling. The spectrum where such an inconsistency happens is the brightest one displayed in Fig.~\ref{eta}. We note that this spectrum also does not show the usual reflection signatures, such as a reflection hump towards higher energies. 

\section{Summary \& Conclusions}
From the analysis of all archival \textit{XMM} data and the newest \textit{Suzaku} data for Mrk\,841 it is evident that the soft X-ray excess is highly variable by a factor of 2 or more. Assuming that an additional Comptonization component is the correct model and \citet{noda:12a} is correct, it is possible to estimate the variability timescale of the excess to larger than 2\,days but less than 7\,days. We discover that the soft excess seems to follow a distinct variability pattern. 

The lack of any clear correlation for the reflection modeling of the soft excess could either be caused by data of insufficient quality to yield sensible constraints and break possible degeneracies or by the assumptions made to construct the model. The approximation of an ionization gradient in an accretion disk by only two ionization zones is only a rough approximation. To be certain that this idea can be excluded a better model will be necessary.
   
However, to learn we need to study more objects to learn whether this is a general property of the soft excess. The origin of soft excess can be further narrowed down by studying simultaneous SEDs and doing broad band fits (e.g. using \textit{Nustar}) as the spectral models differ significantly at hard X-rays.

\bibliographystyle{aa}

\end{document}